# FREQUENCY CHARACTERISTICS OF ACOUSTIC-GRAVITY EIGENWAVES OF THE NONADIABATIC ATMOSPHERE

Lyahov V.V.[*], Neshchadim V.M.

Institute of Ionosphere, 050020, Almaty, Kazakhstan

The main objective of the given work consists in the account of the basic mechanisms, which decline processes in atmosphere from adiabaticity approach, and in the study of effect non-adiabaticity on the swing of eigenmodes of atmosphere at various altitudes. The dispersion equation is solved, dispersion curves for acoustic and gravity branches are explored. The increment of growth of eigenmodes at various altitudes of atmosphere is explored. Frequencies (real part) both for acoustic, and for gravity branches acoustic-gravity waves (AGW) in nonadiabatic atmosphere, coinciding with frequencies of the adiabatic model for the same wave vectors at low altitudes, smoothly decrease with altitude. It leads to decrease of a phase velocity for both branches at larger altitudes in a real atmosphere in comparison with predicted by the adiabatic model.

## Introduction

The given work is continuation of study [1]. Now theoretical models of acoustic-gravity waves (AGW) in an adiabatic atmosphere are widely used. It explained, in particular, by simplicity and obviousness of gained results. In our work we want without excessive detailing to include in study the basic mechanisms which can decline studied processes from adiabaticity approach, and to study effect non adiabaticity on AGW propagation. The analysis of thermal and radiation balances of an atmosphere which has allowed evaluating a degree of importance of the various processes conducting to deviation of a real atmosphere from a requirement adiabaticity is carried out. It has given the chance to include correctly in the basic hydrodynamic equations the corresponding terms modeling most important of these processes, and on this basis to study problems of a stability of an atmosphere and a AGW dissipation in it.

## The short derivation of a dispersion equation

Dynamics of medium, as usually, is described by system of the hydrodynamic equations: Euler's equation, the continuity equation and the energy equation.

$$\rho \frac{\partial \vec{U}}{\partial t} + \rho (\vec{U}\nabla)\vec{U} = -\nabla P + \rho \vec{g}; \quad (1)$$

$$\frac{\partial \rho}{\partial t} = -\nabla(\rho \vec{U}); \quad (2)$$

$$\frac{dP}{dt} - c^2 \frac{d\rho}{dt} = \frac{P}{c_v \rho T}\nabla \vec{J}_c + \frac{P}{c_v \rho T}\nabla \vec{J}_\kappa - \frac{P}{c_v \rho T}\nabla \vec{J}_a - \frac{P}{c_v \rho T}\nabla \vec{L}. \quad (3)$$

All variables depend on a time t and two spatial coordinates: horizontal distance x and altitude z, at that z-axis is upward.

Unlike the adiabatic case, the right member of last equation is nonzero and represents balance of heat for an atmosphere. The first term in a right side of equation (3) describes inflow of heat to an atmosphere at the expense of solar radiation, second - inflow of heat at the expense of condensation of water steams, the third - outflow of energy in the course of atmosphere infrared radiation, the fourth - inflow of heat to an atmosphere, or outflow of heat from an atmosphere to a ground substrata as a result of thermal conductivity. Letter J marks a corresponding energy flow, L - heat flow.

---

[*] Corresponding author. Tel.: + 7 727 3803054; fax: +7 727 3803053
E-mail address: v_lyahov@rambler.ru (V. Lyahov).

After modeling of flows of heat in a right side of equation (3) from system of the hydrodynamic equations the basic equation - the equation for speed of the substance movement is gained:

$$\frac{\partial^2 \vec{U}}{\partial t^2} = \frac{1}{\rho^2} \nabla P(-\rho \nabla \vec{U} - \vec{U} \nabla \rho) - \frac{1}{\rho} \nabla \left[ -c^2 \rho \nabla \vec{U} - (\vec{U} \nabla) P \right] -$$
$$- \frac{1}{\rho} \nabla \left[ \frac{R}{c_v} (\mu \rho J_c) + \frac{R}{c_v} \nabla (L_\kappa \rho_{s0} \exp(-\frac{z}{h}) \vec{U}) - \frac{R}{c_v} \nabla (\sigma T^4 \vec{z}^0) - \right.$$
$$\left. - \frac{R}{c_v} \nabla (-\kappa \nabla T) \right]. \qquad (4)$$

The solution of this equation was searched by a perturbation technique:

$$\vec{U} = \vec{U}_0 + \vec{U}'; P = P_0 + P'; \rho = \rho_0 + \rho'; J_c = J_{c0} + J'_c; T = T_0 + T' \qquad (5)$$

Velocity perturbation has been presented in the form of the sum of stationary and non-stationary parts

$$\vec{U}'(x,z,t) = \vec{V}(x,z,t) + \vec{S}(x,z). \qquad (6)$$

As a result the input equation (4) was divided into two - the equation describing a stationary macroscopic flow and the equation:

$$\frac{\partial^2 \vec{V}}{\partial t^2} = c_0^2 \text{graddiv} \vec{V} - g \nabla V_z - (1-\gamma) \vec{g} \text{div} \vec{V} -$$
$$- \frac{1}{\rho_0} \nabla \left[ \frac{R\mu}{c_v} (\rho' J_{c0} + \rho_0 J'_c) + \frac{RL_\kappa \rho_{s0}}{c_v} \nabla (\exp(-\frac{z}{h}) \vec{V}) - \right.$$
$$\left. - \frac{4\sigma R T_0^3}{c_v} \frac{\partial T'}{\partial z} + \frac{R\kappa}{c_v} \left( \frac{\partial^2 T'}{\partial x^2} + \frac{\partial^2 T'}{\partial z^2} \right) \right], \qquad (7)$$

defining a non-stationary component of motion.

The solution (7) was searched in the form of plane-wave expansion exp (-iωt + k$_x$x + ik$_z$z), and requirement its non trivialness is the dispersion equation:

$$\omega^4 + (M+F)\omega^3 + (H_{2z} + FM + H_{1x} - GL)\omega^2 +$$
$$+ (FH_{2z} + MH_{1x} - GH_{2x} - LH_{1z})\omega + H_{1x}H_{2z} - H_{1z}H_{2x} = 0, \qquad (8)$$

where где

$$F = (ik_x k_z^2 B - k_x k_z A + ik_x^3 B)\left(\frac{1}{Rk_x} + i\frac{k_z T_0}{gk_x}\right) - \frac{\rho_0 k_z D}{g} - \frac{\rho_0 c_v k_z ED}{Rg(ik_z - \mu \rho_0)};$$

$$G = -i\frac{T_0}{g}(ik_x k_z^2 B - k_x k_z A + ik_x^3 B) + \frac{\rho_0 k_x D}{g} + \frac{\rho_0 c_v k_x ED}{Rg(ik_z - \mu \rho_0)};$$

$$L = (ik_x^2 k_z B - k_z^2 A + ik_z^3 B)(\frac{1}{Rk_x} + i\frac{k_z T_0}{gk_x}) + i(\mu\rho_0 D + ik_z D)\frac{\rho_0 k_z}{gk_x} -$$

$$- i\frac{\rho_0 c_v D(\frac{g}{RT_0} E - ik_z E)}{Rg(ik_z - \mu\rho_0)}\frac{k_z}{k_x};$$

$$M = -i\frac{T_0}{g}(ik_x^2 k_z B - k_z^2 A + ik_z^3 B) - i(\mu\rho_0 D + ik_z D)\frac{\rho_0}{g} +$$

$$+ i\frac{\rho_0 c_v D(\frac{g}{RT_0} E - ik_z E)}{Rg(ik_z - \mu\rho_0)}\frac{k_z}{k_x}; \tag{9}$$

$$H_{1x} = -k_x^2(c_0^2 - W);$$

$$H_{1z} = -ik_x g - k_x k_z(c_0^2 - W) + ik_x\frac{W}{h};$$

$$H_{2x} = ik_x(1-\gamma)g - k_x k_z(c_0^2 - W) + ik_x\frac{W}{h};$$

$$H_{2z} = -ik_z\gamma g - k_z^2(c_0^2 - W) + ik_z\frac{2W}{h} - \frac{W}{h^2};$$

The dispersion equation (8) for AGW in a non-adiabatic atmosphere coincides in the limiting case of an adiabatic atmosphere with a known dispersion equation (see, for example, [2]):

$$\omega^4 - c_0^2(k_x^2 + k_z^2)\omega^2 - i\gamma g k_z \omega^2 + (\gamma - 1)g^2 k_x^2 = 0. \tag{10}$$

**Results of the solution of a dispersion equation**

The dispersion equation (8) represents the algebraic equation of the fourth order with the complex coefficients (9) depending on altitude.

The solution of this equation for various atmospheric layers was fulfilled numerically in MAPLE-14 package.

On Fig. 1a, b, c, d, e, f dispersion curves for eigenmodes of a non-adiabatic atmosphere at various altitudes are presented. Horizontal propagation $k_z$=0 was considered. For the real values of a component of a wave vector $k_x$ from a dispersion equation (8) there were 4 complex frequencies $\omega$: two roots with a positive real part and two roots with the negative real part. On figures waves with a positive real part, moving on a *x*-axis are shown only. The negative real part simply corresponds to a wave directed against a *x*-axis.

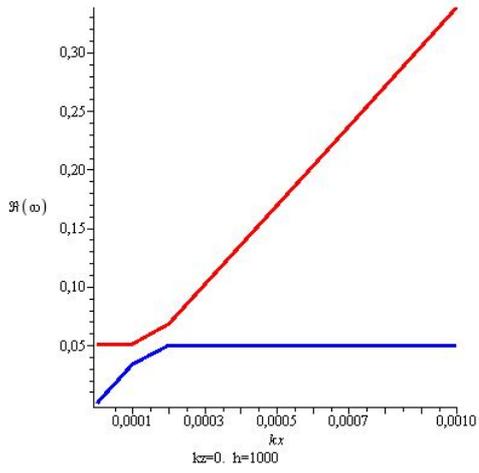

Fig. 1a

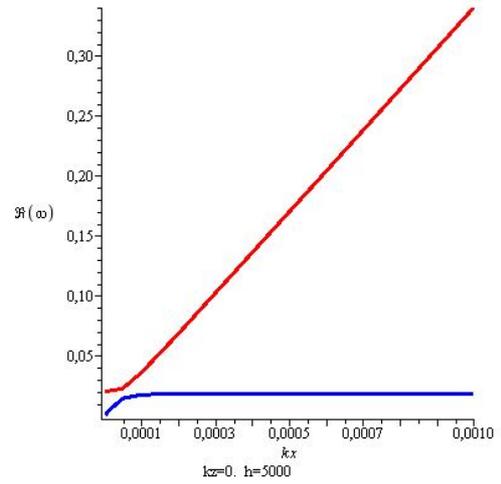

Fig. 1b

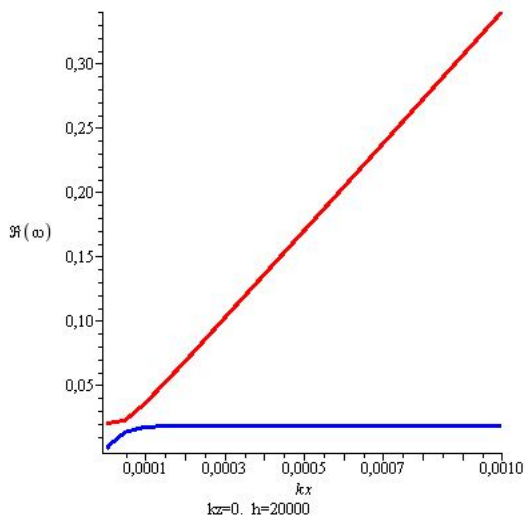

Fig. 1c

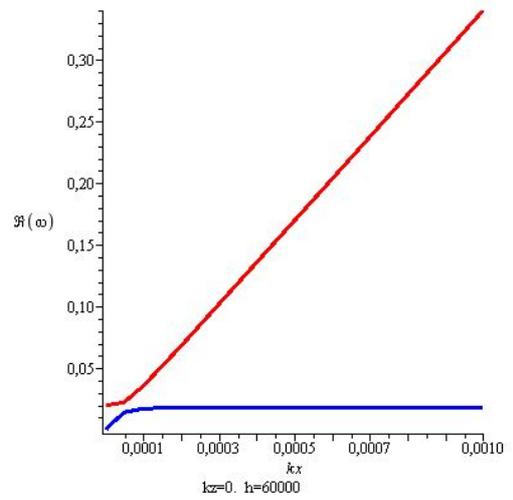

Fig. 1d

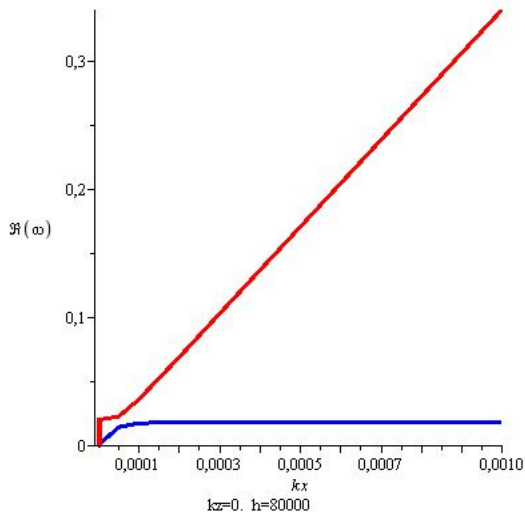

Fig. 1e

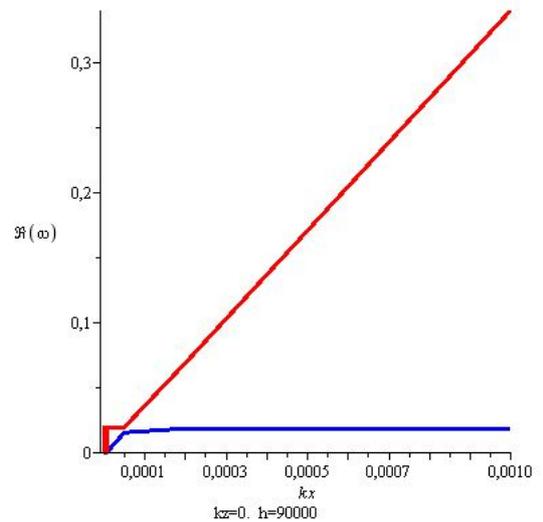

Fig. 1f

As it seen from figures, the dispersion curve, as well as in case of an adiabatic atmosphere, consists of two branches - acoustic (high-frequency) and gravity (low-frequency). However in case of a nonadiabatic atmosphere boundary values of frequencies for acoustic branch $\omega_a = 0.020\,\text{s}^{-1}$ and for gravity branch $\omega_g = 0.018\,\text{s}^{-1}$ coincide with same for an adiabatic atmosphere, beginning only with altitude of 5000 m. At small altitudes to 5000 m these boundary values of frequencies are equal: $\omega_a = 0.051\,\text{s}^{-1}$ and $\omega_g = 0.050\,\text{s}^{-1}$. It is seen that at upper altitudes, since 80000 m, the long-wave acoustic modes miss (a real part of frequency in this field it is equal to zero).

On Fig. 2a, b, c, d, e, f dispersion curves for vertical eigenmodes of a non-adiabatic atmosphere at various altitudes are presented. Vertical propagation $k_x=0$ was considered. On figures waves as with a positive real part of frequency moving upwards and waves with the negative real part moving downwards are shown.

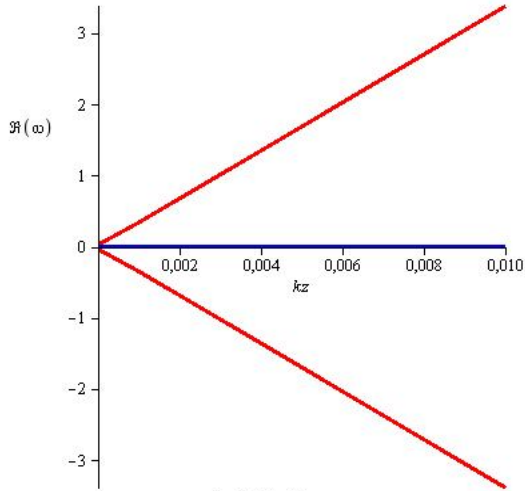
Fig. 2a

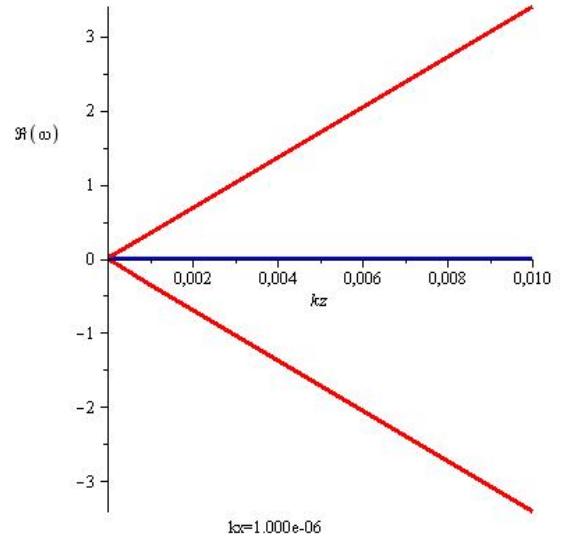
Fig. 2b

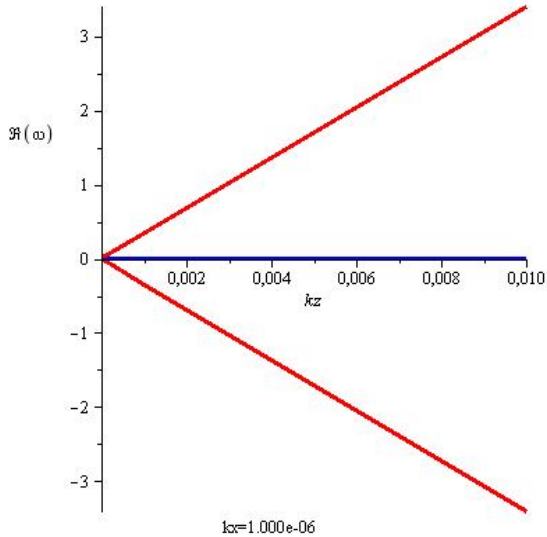
Fig. 2c

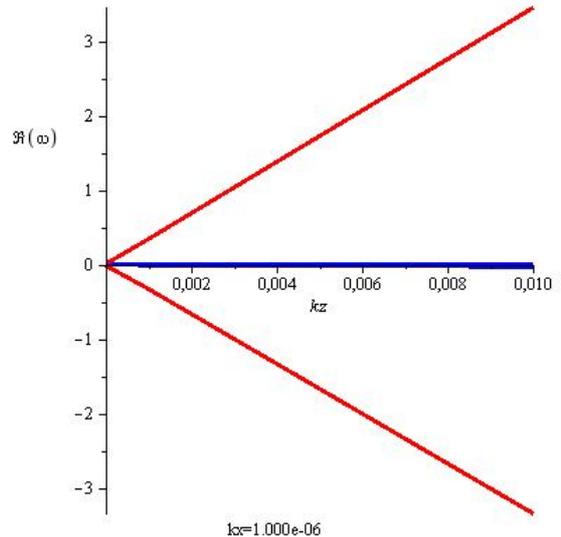
Fig. 2d

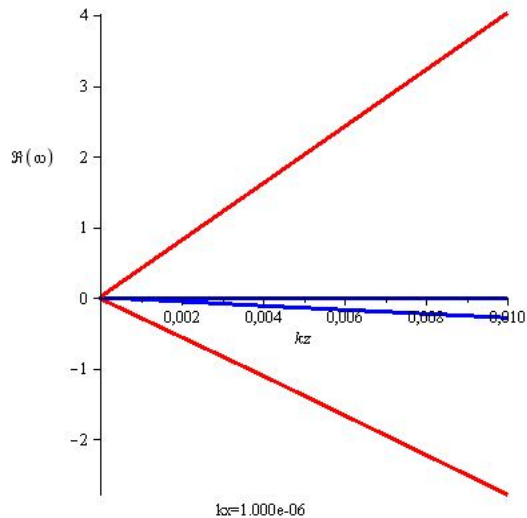
Fig. 2e

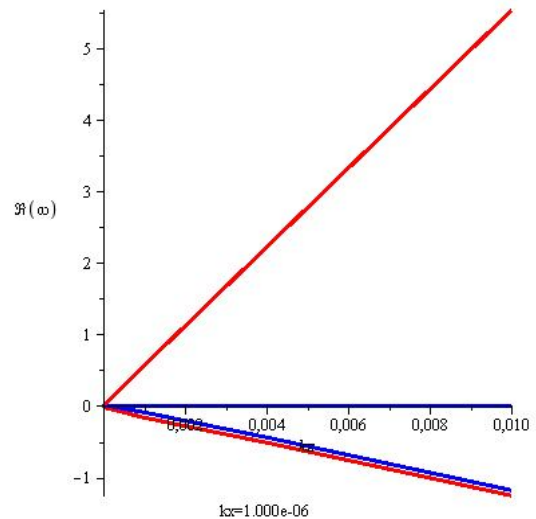
Fig. 2f

On figures it is seen that in this case there are also two branches - high-frequency and low-frequency, merging practically with an abscissa axis. To altitudes of 80000 m dispersion curves are symmetrical. Phase velocity of the waves propagated upwards and downwards coincides and are equal about a sound velocity at the given altitude. But, begin 80000 m symmetry is broken: the phase velocity of the wave propagated upwards becomes more sound velocity, and for a wave propagated downwards - there is less than sound velocity. Moreover, the low-frequency branch at altitude of 90000 m for waves, moving downwards, coincides with a high-frequency branch.

In tables 1a, b and 2a, b signs of imaginary parts of eigenmodes at various altitudes for horizontal and vertical propagating accordingly are given.

Table 1a. Horizontal propagation $k_z=0$ (gravity branch)

| h(m) | $k_x(m^{-1})$ | $1 \cdot 10^{-6}$ | $5 \cdot 10^{-6}$ | $1 \cdot 10^{-5}$ | $5 \cdot 10^{-5}$ | $1 \cdot 10^{-4}$ | $2 \cdot 10^{-4}$ | $6 \cdot 10^{-4}$ | $1 \cdot 10^{-3}$ |
|---|---|---|---|---|---|---|---|---|---|
| 1000 | | - | - | - | - | - | + | + | + |
| 5000 | | + | + | + | + | + | + | + | + |
| 20000 | | - | - | - | - | - | - | - | - |
| 40000 | | - | - | - | - | + | + | + | + |
| 60000 | | - | - | - | - | + | + | + | + |
| 80000 | | - | - | - | - | + | + | + | + |
| 90000 | | - | - | - | - | + | + | + | + |

Table 1b. Horizontal propagation $k_z=0$ (acoustic branch)

| h(m) | $k_x(m^{-1})$ | $1 \cdot 10^{-6}$ | $5 \cdot 10^{-6}$ | $1 \cdot 10^{-5}$ | $5 \cdot 10^{-5}$ | $1 \cdot 10^{-4}$ | $2 \cdot 10^{-4}$ | $6 \cdot 10^{-4}$ | $1 \cdot 10^{-3}$ |
|---|---|---|---|---|---|---|---|---|---|
| 1000 | | + | + | + | + | + | - | - | - |
| 5000 | | + | + | + | - | - | - | - | - |
| 20000 | | - | - | - | - | - | - | - | - |
| 40000 | | - | - | - | - | - | - | - | - |
| 60000 | | - | - | - | - | - | - | - | - |
| 80000 | | - | - | - | - | - | - | - | - |
| 90000 | | - | - | - | - | - | - | - | - |

Table 2a. Vertical propagation $k_x=0$ (gravity branch)

| h(m) | $k_z(m^{-1})$ | $1 \cdot 10^{-6}$ | $5 \cdot 10^{-6}$ | $1 \cdot 10^{-5}$ | $1 \cdot 10^{-4}$ | $1 \cdot 10^{-3}$ | $4 \cdot 10^{-3}$ | $1 \cdot 10^{-2}$ |
|---|---|---|---|---|---|---|---|---|

| | | | | | | | |
|---|---|---|---|---|---|---|---|
| 1000 | - | + | + | + | + | + | + |
| 5000 | + | + | + | + | + | + | + |
| 20000 | - | - | - | + | + | + | + |
| 40000 | - | - | - | + | + | + | + |
| 60000 | - | - | - | + | + | + | + |
| 80000 | - | - | - | + | + | + | + |
| 90000 | - | - | - | + | + | + | + |

Table 2b. Vertical propagation $k_x = 0$ (acoustic branch)

| h(m) | $k_z(m^{-1})$ | $1 \cdot 10^{-6}$ | $5 \cdot 10^{-6}$ | $1 \cdot 10^{-5}$ | $1 \cdot 10^{-4}$ | $1 \cdot 10^{-3}$ | $4 \cdot 10^{-3}$ | $1 \cdot 10^{-2}$ |
|---|---|---|---|---|---|---|---|---|
| 1000 | | - | - | - | - | - | - | - |
| 5000 | | + | - | - | - | - | - | - |
| 20000 | | - | - | - | - | - | - | - |
| 40000 | | - | - | - | - | - | - | - |
| 60000 | | - | - | - | - | - | - | - |
| 80000 | | - | - | - | - | - | - | - |
| 90000 | | - | - | - | - | - | - | - |

**Conclusion**

In an adiabatic atmosphere at the real wave vectors the computed frequencies also are real that indicates to absence of AGW dissipation. In a non-adiabatic model of atmosphere, more adequate reality, frequencies are complex, and in some atmosphere layers the imaginary part is negative that testifies about damping of waves generated at corresponding altitudes; in other layers of imaginary parts are positive that speaks about instability of atmosphere to corresponding modes at these altitudes.

As it seen from Table 1a at horizontal propagating gravity waves are swing more intensively at small lengths of waves and at upper altitudes. Acoustic waves (see Table 1c), on the contrary, are more intensively swing at larger lengths of waves and at low altitudes. At vertical propagation (see Table 2a) gravity waves are swing at small lengths of waves, and acoustic waves (see Table 2c) damp practically at all lengths of waves and at all altitudes.

Frequencies (real part) both for acoustic, and for gravity branches AGW in nonadiabatic atmosphere, coinciding with frequencies of the adiabatic model for the same wave vectors at low altitudes, smoothly decrease with altitude. It leads to decrease of a phase velocity for both branches at larger altitudes in a real atmosphere in comparison with predicted by the adiabatic model. Last conclusion seems physically reasonable as a phase velocity of acoustic and gravity modes are tied up to a sound velocity in medium and it decreases together with vertical fall of real temperature of atmosphere.

Value of the relation of an imaginary part to real part of frequency, small at low altitudes, is increased with altitude growth, reaching almost unity for some modes of gravity waves at altitude of mesopause.